\documentclass[aps,prb,twocolumn,floats,showpacs,floatfix,a4paper]{revtex4}
\usepackage{epsfig}
\usepackage{color}
\usepackage{bm}
\usepackage{latexsym}

\def\mps{M_{\rm PS}(t)}

\begin{document}
\newcommand{\cc}{{\bf\Large C }}
\newcommand{\hide}[1]{}
\newcommand{\tbox}[1]{\mbox{\tiny #1}}
\newcommand{\half}{\mbox{\small $\frac{1}{2}$}}
\newcommand{\sinc}{\mbox{sinc}}
\newcommand{\const}{\mbox{const}}
\newcommand{\trc}{\mbox{trace}}
\newcommand{\intt}{\int\!\!\!\!\int }
\newcommand{\ointt}{\int\!\!\!\!\int\!\!\!\!\!\circ\ }
\newcommand{\eexp}{\mbox{e}^}
\newcommand{\EPS} {\mbox{\LARGE $\epsilon$}}
\newcommand{\ar}{\mathsf r}
\newcommand{\im}{\mbox{Im}}
\newcommand{\re}{\mbox{Re}}
\newcommand{\bmsf}[1]{\bm{\mathsf{#1}}}
\newcommand{\dd}[1]{\:\mbox{d}#1}
\newcommand{\abs}[1]{\left|#1\right|}
\newcommand{\bra}[1]{\left\langle #1\right|}
\newcommand{\ket}[1]{\left|#1\right\rangle }
\newcommand{\mbf}[1]{{\mathbf #1}}
\newcommand{\eos}{\,.}
\definecolor{red}{rgb}{1,0.0,0.0}

\title{Wavepacket dynamics of the nonlinear Harper model}
\author{Gim Seng Ng and Tsampikos Kottos}

\affiliation{
Department of Physics, Wesleyan University, Middletown, Connecticut 06459, USA 
}

\begin{abstract}
The destruction of anomalous diffusion of the Harper model at criticality, due to weak nonlinearity $\chi$, is analyzed. 
It is shown that the second moment grows subdiffusively as $\langle m_2\rangle \sim t^{\alpha}$ up to time $t^*\sim \chi^{\gamma}$. 
The exponents $\alpha$ and $\gamma$ reflect the multifractal properties of the spectra and the eigenfunctions of the linear model. 
For $t>t^*$, the anomalous diffusion law is recovered, although the evolving profile has a different shape than in the linear case. 
These results are applicable in wave propagation through nonlinear waveguide arrays and transport of Bose-Einstein condensates in optical lattices.
\\
\end{abstract}

\maketitle


\section{Introduction}
The spreading of a quantum mechanical wavepacket for a particle moving in a periodic lattice is a textbook example. 
One finds that the width increases ballistically with time and the corresponding eigenstates are extended (Bloch states). 
In the opposite case of strongly disordered systems, the eigenstates are exponentially localized, resulting in a total halt of the wavepacket spreading in the long-time limit \cite{A58}. 
Between these extremes flourishes the world of quantum systems with anomalous diffusion.
These include systems studied in the early days of quantum mechanics, such as Bloch electrons in a magnetic field \cite{P33} as well as quasicrystals \cite{SBGC84} and disordered systems at the metal-insulator transition \cite{MK93}. 
In these cases, the eigenfunctions (or even the spectrum) show a fractal
structure which dictates the wavepacket spreading.

The physical motivation to study such systems, along with the
mathematically intriguing nature of their spectra, led to a series of works
that eventually advanced our understanding of their dynamical properties. 
Most of these works have focused on the analysis of the temporal decay of
the survival probability $P(t)$ at the initial position $n_0$ and 
the growth of the wavepacket second moment $m_2(t)$ \cite{G93,EK93,KKKG97,ZDSORPZN01}. 

Despite all this activity, nothing is known about the effect of nonlinearity in the wavepacket spreading for systems with fractal spectra and eigenfunctions.
Noticeable exceptions are Refs.~9 and 10
 which numerically studied the wavepacket dynamics for the prototype one-dimensional (1D)
tight-binding Harper model with nonlinearity. 
Their conclusions, however,  are contradictory. 
Although both studies conclude that {\it infinitesimal} nonlinearity results
in a short-time subdiffusive spreading of the variance $m_2\sim t^{\alpha}$,
they give different values for the power-law exponent. 
While Ref.~10
 concludes that the subdiffusive spreading persists for
longer time, Ref.~9
 reports a saturation of the second moment.
Moreover, they both fail to identify traces of the underlying linear model's multifractality in the dynamics generated by the corresponding nonlinear one. 

In the present paper, we address the nonlinearity-induced destruction of the anomalous diffusion. 
Our focus is on the 1D Harper model at criticality but we expect that our results will be applicable for other critical models (such as Fibonacci lattices) as well. 
Our detailed numerical calculations and theoretical arguments show that there is a {\it critical nonlinearity} $\chi^*$, above which the nature of anomalous diffusion (associated with the linear model) is altered. 
In particular, we found that for parametrically large time, the average (over initial sites) spreading of an initially localized $\delta-$like wavepacket spreads as
\begin{equation}
\label{nlvar}
\langle m_2(t) \rangle 
\propto t^{\alpha}\quad {\rm with }\quad 
2\beta -D_2^{\mu}\leq \alpha\leq 2\beta
\end{equation}
where $D_2^{\mu}$ is the the fractal dimension of the Local 
Density of States (LDoS) and $\alpha$ is {\it independent} of the nonlinearity $\chi$. The time scale $t^*$ up to 
which Eq.~(\ref{nlvar}) is observed depends on $\chi$ as 
\begin{equation}
\label{nltau}
t^* \sim \chi^{\gamma}\quad {\rm with} \quad \gamma= 1/D_2^{\mu}
\end{equation}
Beyond this time scale, the effects of nonlinearity diminish and we recover the spreading 
of the linear model i.e. $\langle m_2(t) \rangle \propto t^{2\beta}$. However, other moments
still behave differently with respect to the linear (i.e. $\chi =0$) case. Finally, we show that for 
any $\chi\neq 0$, the core of the evolving profile satisfies the scaling behavior,
\begin{equation}
\label{scas}
P_{\rm s}(x,t) = \sqrt{\langle m_2(t) \rangle} P_{\rm \chi}(n,t),\quad x\equiv
{n-n_0\over\sqrt{\langle m_2(t) \rangle}}
\end{equation}

This paper is structured in the following way. In Sec. II an overview of the linear Harper model is presented. The nonlinear Harper model will be introduced in Sec. III and the effects of nonlinearity on the spreading and the evolving profile will be investigated both numerically and analytically. Finally, we will draw our conclusions in Sec. IV.
\section{The Harper Model: Overview}
In this section, we will briefly review the basic properties of the linear Harper Model.
Its dynamics is described by the standard 1D tight-binding model, 
\begin{equation}
\label{dnlslinear}
i{\partial \psi_n(t)\over \partial t} = \psi_{n+1}(t)+\psi_{n-1}(t) +
V_n\psi_n(t) \,\, 
\end{equation}
Above, $\psi_n(t)$ denotes the probability amplitude for a particle to be at site $n$ at time $t$ and we will assume that the on-site
potential $V_n$ takes the form:
\begin{equation}
\label{vn}
V_n=\lambda \cos (2\pi\sigma n + \phi) 
\end{equation}
When $\sigma$ is an irrational number, the period 
of the on-site potential 
$V_n$ is incommensurate with the lattice period. In such cases, the eigenstates 
of the linear system are extended when $\lambda<2$ and the spectrum consists of bands (ballistic 
regime). For $\lambda>2$, the spectrum is point-like and all states are exponentially localized 
(localized regime). The most interesting case is the critical point $\lambda=2$, where we have 
a metal-insulator transition. At this point, the spectrum is a zero measure Cantor set 
\cite{frank}, while the eigenstates show self-similar fluctuations on all scales \cite{GKP95,AA80}.

For the critical case, where $\lambda=2$, previous studies show that the temporal decay of the probability to remain at the original site $n_0$ up to time $t$ goes as
\begin{equation}
\label{plinear}
P(t) \equiv |\psi_{n_0}(t)|^2 \sim 1/t^{D^{\mu}_2}
\end{equation}
while the wavepacket second moment grows as 
\begin{equation}
 m_2(t) \equiv \sum_n (n-n_0)^2  |\psi_{n}(t)|^2 \sim t^{2\beta} \quad {\rm with }\quad 
\beta\geq D_2^{\mu}/D_2^{\psi}
\end{equation}
where the exponent $\beta$  depends on both $D_2^{\mu}$ and the correlation dimension of the fractal eigenfunctions $D_2^{\psi}.$\cite{KKKG97} 
In fact, recent studies were able to demonstrate that the center 
of the wavepacket spatially decreases as $|n-n_0|^{D_2^{\psi}-1}$ for $|n-n_0|\ll t^{\beta}$ 
while the front shape scales as \cite{ZDSORPZN01}
\begin{equation}
P(n,t)= A(t) \exp(-|(n-n_0)/{\sqrt m_2}|^{1/(1-\beta)})
\end{equation}
where $A(t)$ is the height of the wave profile at time $t$.

\section{The Nonlinear Harper Model}
Introducing an on-site nonlinear term into Eq.~({\ref{dnlslinear}}), we come out with the nonlinear Harper model (NLHM). Mathematically, it is described by the following discrete nonlinear Schr\"odinger (DNLS) equation,
\begin{equation}
\label{dnls}
i{\partial \psi_n(t)\over \partial t} = \psi_{n+1}(t)+\psi_{n-1}(t) + 
V_n \psi_n(t) - \chi |\psi_n(t)|^2 \psi_n(t) \,\,
\end{equation}

We want to investigate the temporal behavior of the variance $\langle m_2(t)\rangle_{\phi}$ for the NLHM at the critical point $\lambda=2$. The initial condition 
is always taken to be a $\delta$-like localized state i.e. $\psi_n(t=0)= \delta_{n,n_0}$. An 
averaging $\langle \cdot\rangle_{\phi}$ over different initial positions $n_0$ (at least $50$) 
of the $\delta$ function [or equivalently over a random distribution of the phase $\phi$ in 
Eq.~(\ref{vn})] is done. In our calculations, we assume $\sigma$ to be equal
to the golden 
mean $\sigma_G=({\sqrt 5}-1)/2$. Equation~(\ref{dnls}) has been integrated numerically using 
a finite-time-step fourth order Runge-Kutta algorithm on a self-expanding lattice in order 
to eliminate finite-size effects \cite{IKPT97}. Whenever the probability of finding the particle 
at the edges of the chain exceeded $10^{-10}$, ten new sites were added to each edge. Numerical 
precision is checked by monitoring the conservation 
of probability (norm) $\sum_n |\psi_n(t)|^2 =1$. In all cases, the deviation from unity was less than $10^{-6}$. 


\subsection{Wavepacket Spreading}
\begin{figure}
\includegraphics[width=\columnwidth,keepaspectratio,clip]{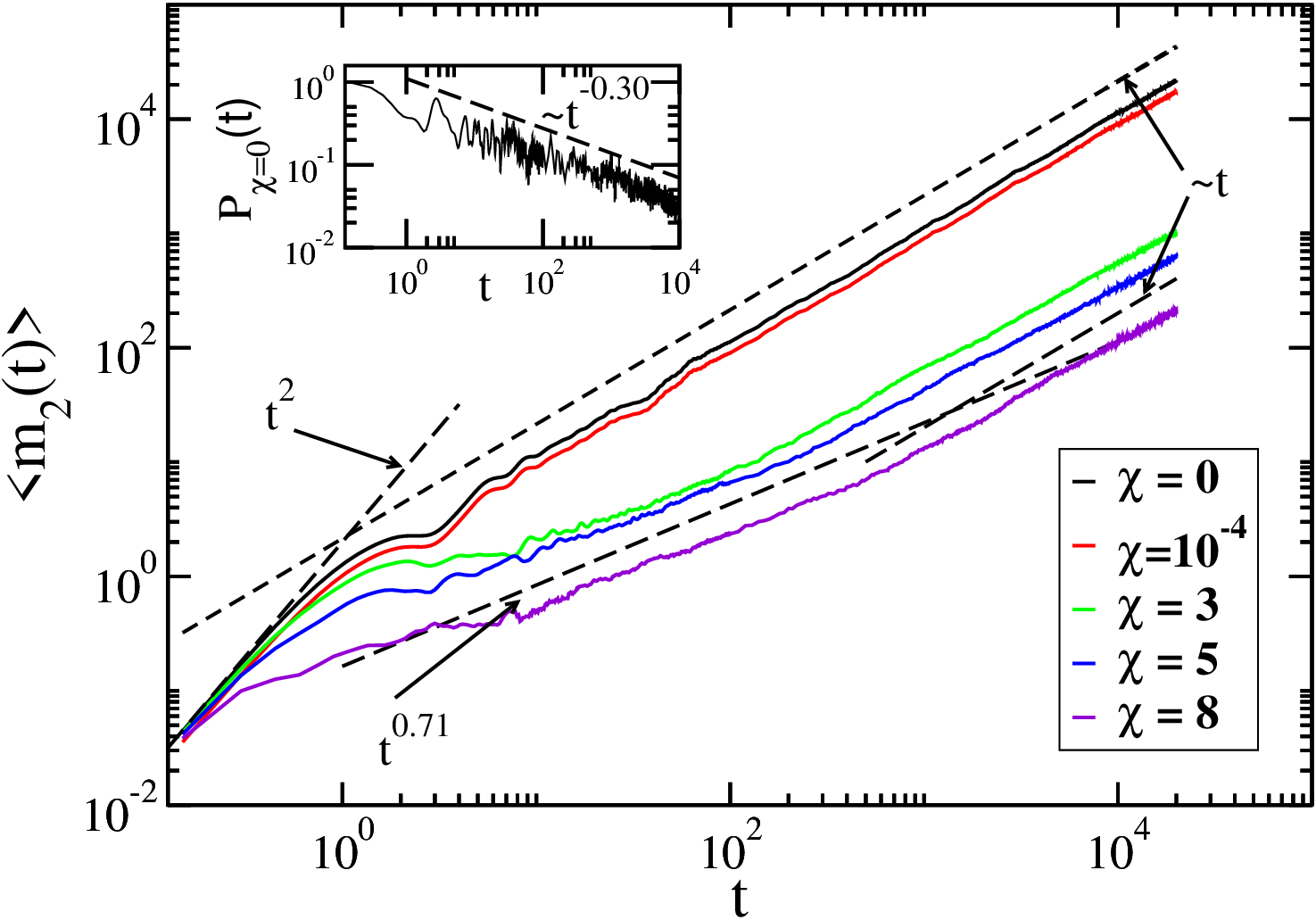}
\caption{\label{cap:fig1}(color online)
The variance $\langle m_2(t)\rangle_{\phi}$ for various $\chi$ values of the NLHM. The 
$\chi=10^{-4}$ case is shifted downwards in order to distinguished it from the $\chi=0$ 
case. Dashed lines have slopes as indicated in the figure and are drawn to guide the eye. 
Inset: the decay of $P(t)\sim 1/t^{D_2^{\mu}}$ for the linear Harper model ($\chi=0$). 
}\end{figure}

In Fig.~\ref{cap:fig1}, we report our numerical results for some representative $\chi$ values in a double 
logarithmic plot. In all cases, the variance displays a power-law behavior $\langle m_2(t)\rangle 
\sim t^{\alpha}$. Leaving aside the very initial spreading $t\leq 1$ (which in all cases 
is ballistic), we observe that for $\chi\geq \chi^*\sim 5.5 \times 10^{-4}$ the spreading is subdiffusive 
with $\alpha<2\beta$. This spreading is valid up to time $t<t^*$ and then relaxes to the anomalous 
diffusion $\alpha=2\beta$ that characterizes the linear Harper model.

To obtain the law of spreading of Eq.~(\ref{nlvar}) for $t<t^*$, we consider the following simple 
model: The initial site together with the nearby sites $n=n_0\pm \delta n$ is considered as a confined source \cite{note1}. Anything emitted from them moves 
according to the spreading law of the linear Harper i.e. $s(t)=\sqrt{\langle m_2(t)\rangle_{\phi}}=v 
t^{\beta}$ since for $t<t^*$ the leaking norm is small and therefore the nonlinear term 
in Eq.~(\ref{dnls}) is negligible with respect to the on-site potential $V_n$.

Initially, all probability is concentrated at the source. Due to the nonlinear nature of the system, 
the decay rate is not constant but depends on the remaining norm. Thus, the decay process is
characterized by the nonlinear equation,
\begin{equation}
 dP(t)/dt = -\Gamma_{P(t)} P(t)
\end{equation}
which leads to non-exponential decay.\cite{SP06} On the other hand, for $\chi=0$, the decay is power-law
\begin{equation}
\label{eqstay}
P(t) \sim 1/t^{\delta}
\end{equation}
with $\delta=D_2^{\mu}$ [see Eq.~(\ref{plinear}) above]. It is, therefore, natural to expect that for $\chi\neq 0$, we will 
have a slower decay with $\delta\leq D_2^{\mu}$ due to self-trapping.\cite{KC85,BDF93,T94} 
The variance of the confined-source model is then given by \cite{HKKG01}
\begin{equation}\label{eqmps}
\mps \approx \int_0^\infty\!\!\!\!\!ds~s^2\int_0^t\!\!\!dt^{'}
(-\dot{P}(t^{'}))\:\delta(s-v(t-t^{'})^{\beta}),\quad
\end{equation}
where $-\dot{P}(t)$ is the flux emitted from the confined source.
Substituting Eq.~(\ref{eqstay}) for $P(t)$ we get
\begin{equation}\label{eqmpspoft}
\mps \sim  v^2 \int_0^t dt^{'} ({1\over t^{'}})^{\delta+1} (t-t^{'})^{2\beta}.\quad
\end{equation}
%
%
If we do a variable substitution $\tilde{t}=t^{'}/t$, we will get
\begin{equation}\label{eqmpspoft2}
\mps \sim v^2~t^{2\beta-\delta} \int_0^1 d\tilde{t}~(1/\tilde{t})^{\delta+1}(1-\tilde{t})^{2\beta}, \quad
\end{equation}
which leads to Eq.~({\ref{nlvar}}) with $\alpha = 2\beta-\delta \geq 2\beta-D_2^{\mu}$.

 In order to compare the results of the numerical simulations
with the theoretical predictions of Eq.~(\ref{nlvar}) we have calculated for the linear Harper
model the decay exponent $\langle D_2^{\mu}\rangle_{\phi}$ of the survival probability (see 
inset of Fig.~\ref{cap:fig1}) and the power-law exponent of the variance $\langle m_2(t)\rangle_{\phi}$. 
The least-squares fit gives the values $\langle D_2^{\mu}\rangle_{\phi} \approx 0.30\pm 
0.03$ and $2\beta \approx 1.00\pm 0.03$. The numerically extracted value of $\alpha=0.71$ 
fulfills nicely the bound $2\beta - D_2^{\mu} = 0.70$ given by Eq.~(\ref{nlvar}).

\begin{figure}
\includegraphics[width=\columnwidth,keepaspectratio,clip]{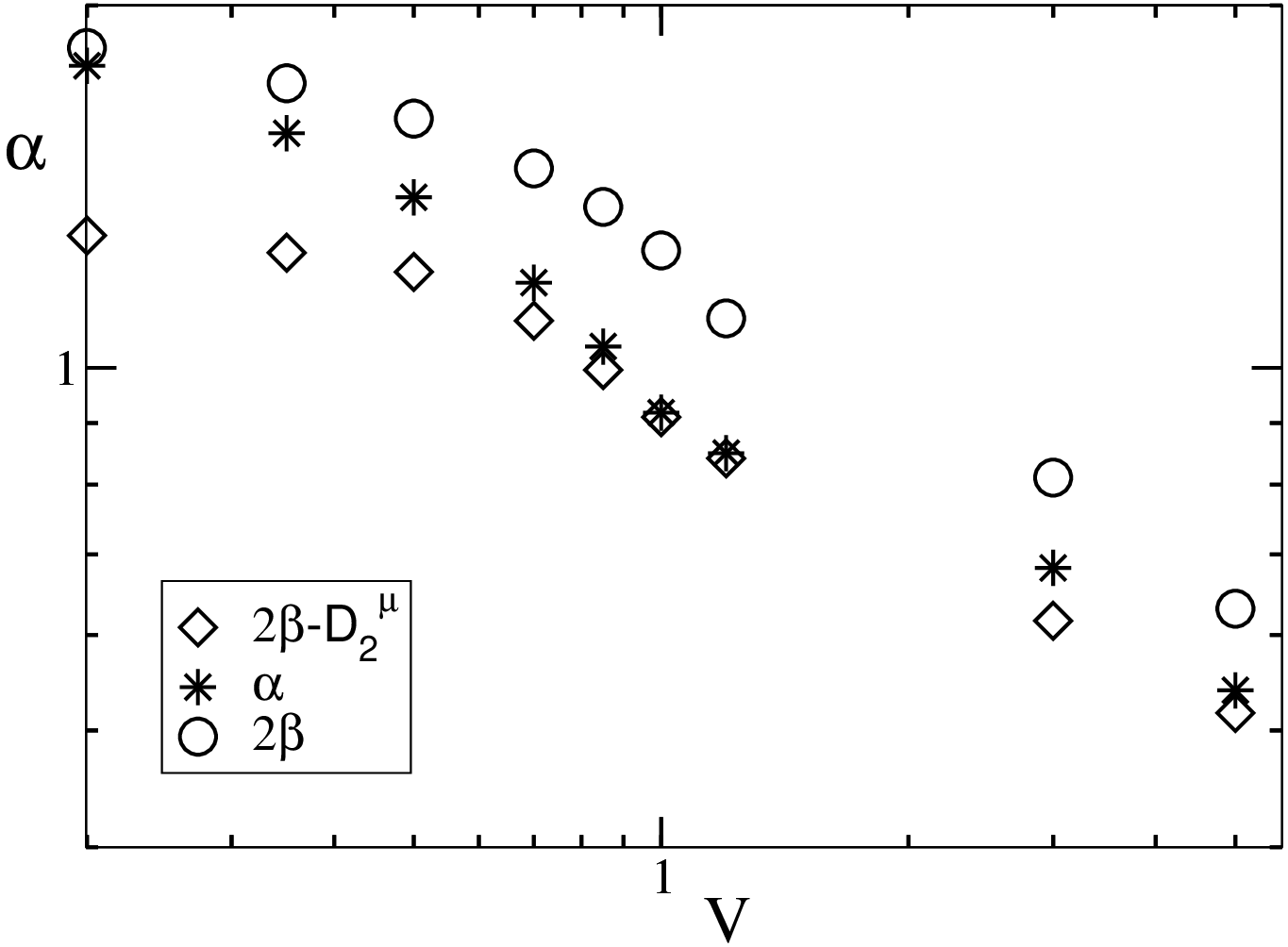}
\caption{\label{cap:fig2}
The fitting values (stars) of the power-law exponent $\alpha$ vs. V for the 
Fibonacci model. Circles are the fitting exponent $2\beta$ of the linear model for the 
spreading $\langle m_2(t) \rangle \sim t^{2\beta}$, while diamonds are the extracted exponents $2\beta-
D_2^{\mu}$, where $D_2^{\mu}$ was obtained from the decay of $P(t)\sim1/t^{D_2^{\mu}}$ of the linear model.}\end{figure}
\begin{figure}
\includegraphics[width=\columnwidth,keepaspectratio,clip]{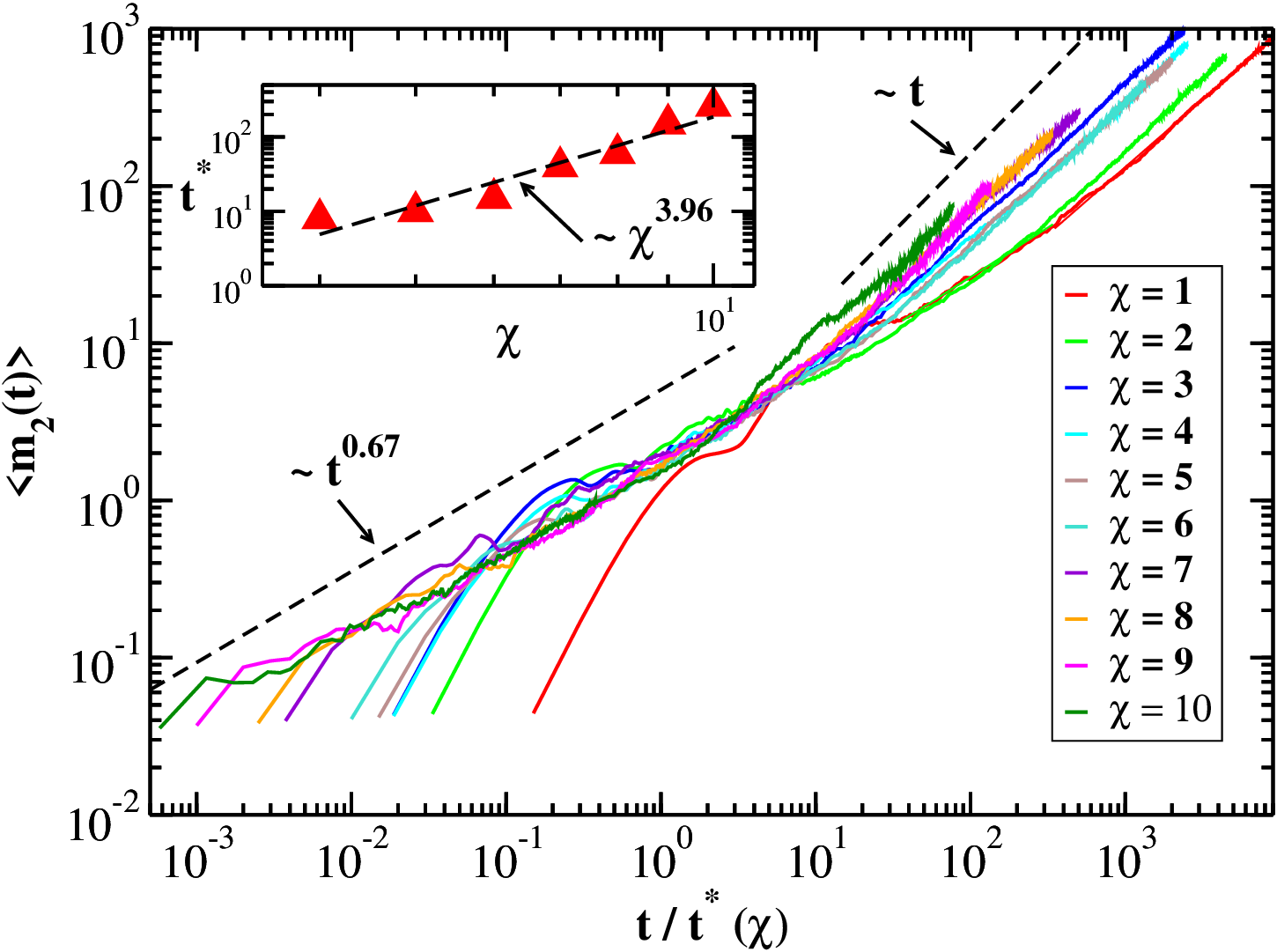}
\caption{\label{cap:fig3}(color online)
The variance vs. scaled time. $t^*$ is extracted by scaling the time-axis so that the variance curves overlap in the time-region where Eq.(\ref{nlvar}) is valid. Inset: the scaling of $t^*$ versus $\chi$. The dashed line indicates the least square fit with slope $3.96$ [the theoretical prediction Eq. (\ref{nltau}) estimates it to be $3.45\pm0.35$].
}\end{figure}

To further test the validity of Eq.~(\ref{nlvar}), we have also performed simulations
with the Fibonacci model, where $D_2^{\psi}$ and $D_2^{\mu}$ (and, therefore, $\beta$) can be 
varied according to the on-site potential $V_n$. The latter takes only two values $\pm V$ 
($V\neq 0$) that are arranged in a Fibonacci sequence.\cite{SBGC84} Again, we find a power-law spreading $\langle m_2(t)\rangle \sim t^{\alpha}$. The extracted exponents $\alpha$ corresponding to
various $V$'s are reported together with the theoretical predictions in Fig.~\ref{cap:fig2} and they confirm the validity of Eq.~(\ref{nlvar}).

From Fig.~\ref{cap:fig1} we see that the time scale $t^*$ up to which the power law of Eq.~(\ref{nlvar}) applies depends on the strength of the nonlinearity parameter $\chi$. 
One can estimate the time $t^*$ from the fact that the effective potential $V_{\chi}= 
-\chi |\psi_{n_0}(t)|^2$ is comparable with the on-site potential $V_{n_0}\sim \lambda$ of the linear
model at $t^*$. After this time, one expects that the effect of nonlinearity on the wavepacket spreading 
is negligible and, therefore, the survival probability should decay as $P(t)= |\psi_{n_0}(t)|^2
\sim 1/t^{D_2^{\mu}}$. Following this line of argumentation,
we have that
\begin{equation}
\label{taustar}
 V_{n_0} \sim \chi|\psi_{n_0}(t^*)|^2 \rightarrow V_{n_0} \sim \chi (1/t^*)^{D_2^{\mu}},
\end{equation}
leading to Eq. (\ref{nltau}). 
To test this theoretical prediction, we have manually scaled the time axis
so that the variance curve where Eq. (\ref{nlvar}) applies overlaps for various $\chi$ values. The extracted scaling parameters $t^*(\chi)$ are plotted in the inset of 
Fig.~\ref{cap:fig3}. One can clearly see that the numerical data confirms the theoretical prediction of Eq.~(\ref{nltau}).

\begin{figure}
\includegraphics[width=\columnwidth,keepaspectratio,clip]{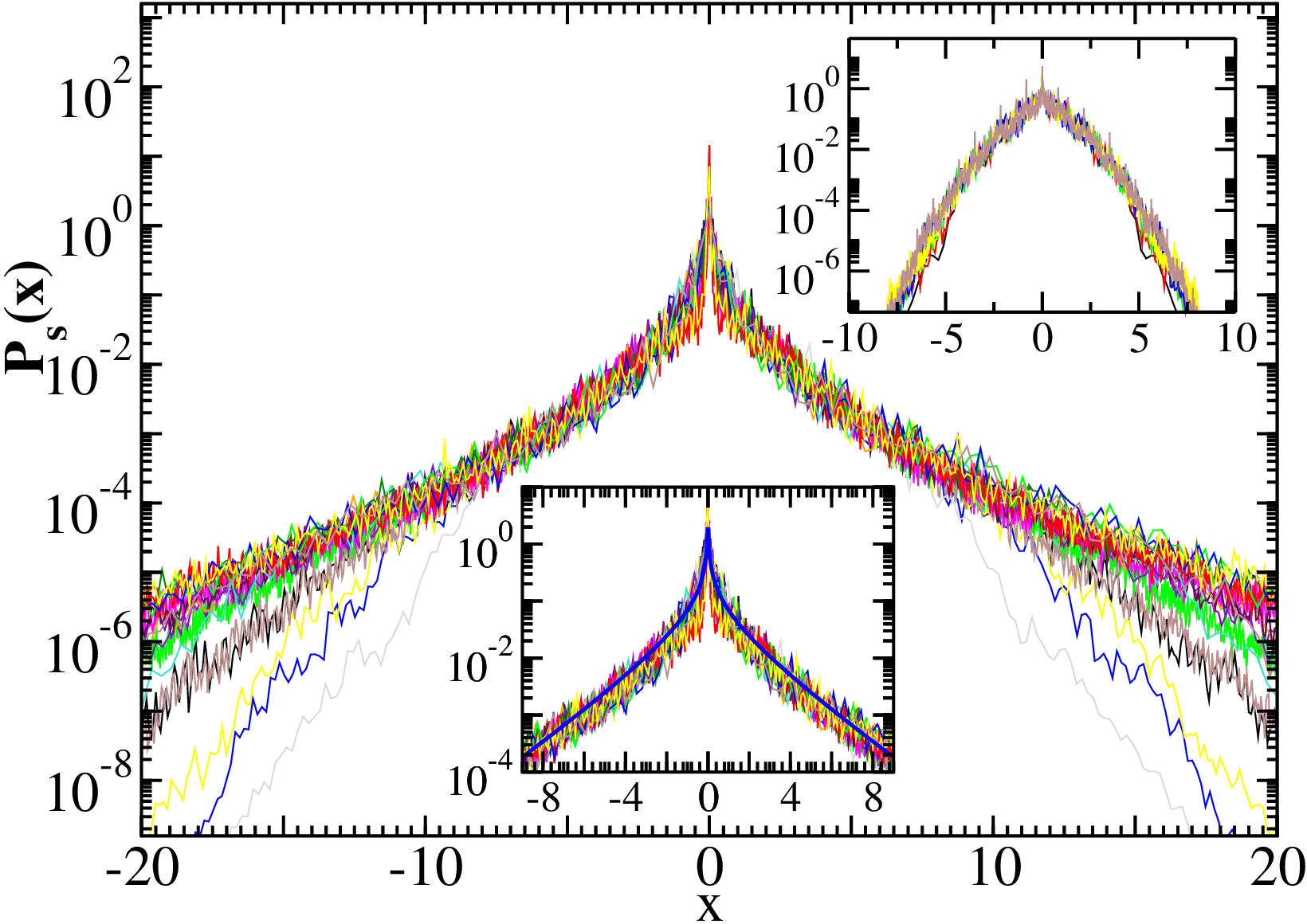}
\caption{\label{cap:fig5}(color online)
The scaled probability distributions $P_s(x)$ versus $x$ of the NLHM at different time
$t$ (including $t>t^*$) and for various $\chi>\chi^*$. The nice overlap of the curves at 
the core $|x|\leq 8$ confirms the validity of the scaling law Eq.~(\ref{scas}). The scaling
is lost for $\abs{x}> 8$. The lower inset shows the behavior near the origin. 
The blue solid line
is the fitting curve Eq.~(\ref{core}). For comparison we also report in the upper inset the 
probability distribution $P_s(x)$ of the linear Harper model. 
}\end{figure}

Finally, we will discuss the shape of the evolving wavefunction for $\chi \geq \chi^*$. Although 
for $t>t^*$ the temporal behavior of the variance becomes the same as that of the linear 
Harper model, other moments differ. The profiles reported 
in Fig.~\ref{cap:fig5} are snapshots of the average wavefunction at various times and for three representative 
values of $\chi=1,3$ and $5$ (the linear case $\chi=0$ is also plotted in the upper inset for
comparison). They are plotted with the scaling assumption in Eq. (\ref{scas}). The data 
collapse for $\abs{x}\leq 8$ (different curves corresponding to various times $t$ and nonlinearity 
strengths $\chi$) reveals that this representation is not affected much by finite-$\chi$ 
corrections, although the tails of the profile are $\chi$-dependent. In fact, we found 
that the probability distribution near the center can be described by the formula:
\begin{equation}
\label{core}
P_s(x)\sim |x|^{-\gamma}\exp(-|x|/l_{\infty}),\,\,\, |x| \leq 8,
\end{equation}
where $\gamma\approx 0.7$ and $l_{\infty}\approx 1.82$. We note that a similar expression for
the core of the probability distribution applies for 1D and quasi-1D disordered models with zero nonlinearity. \cite{G76,IKPT97}

\subsection{Critical Nonlinearity}
\begin{figure}
\includegraphics[width=\columnwidth,keepaspectratio,clip]{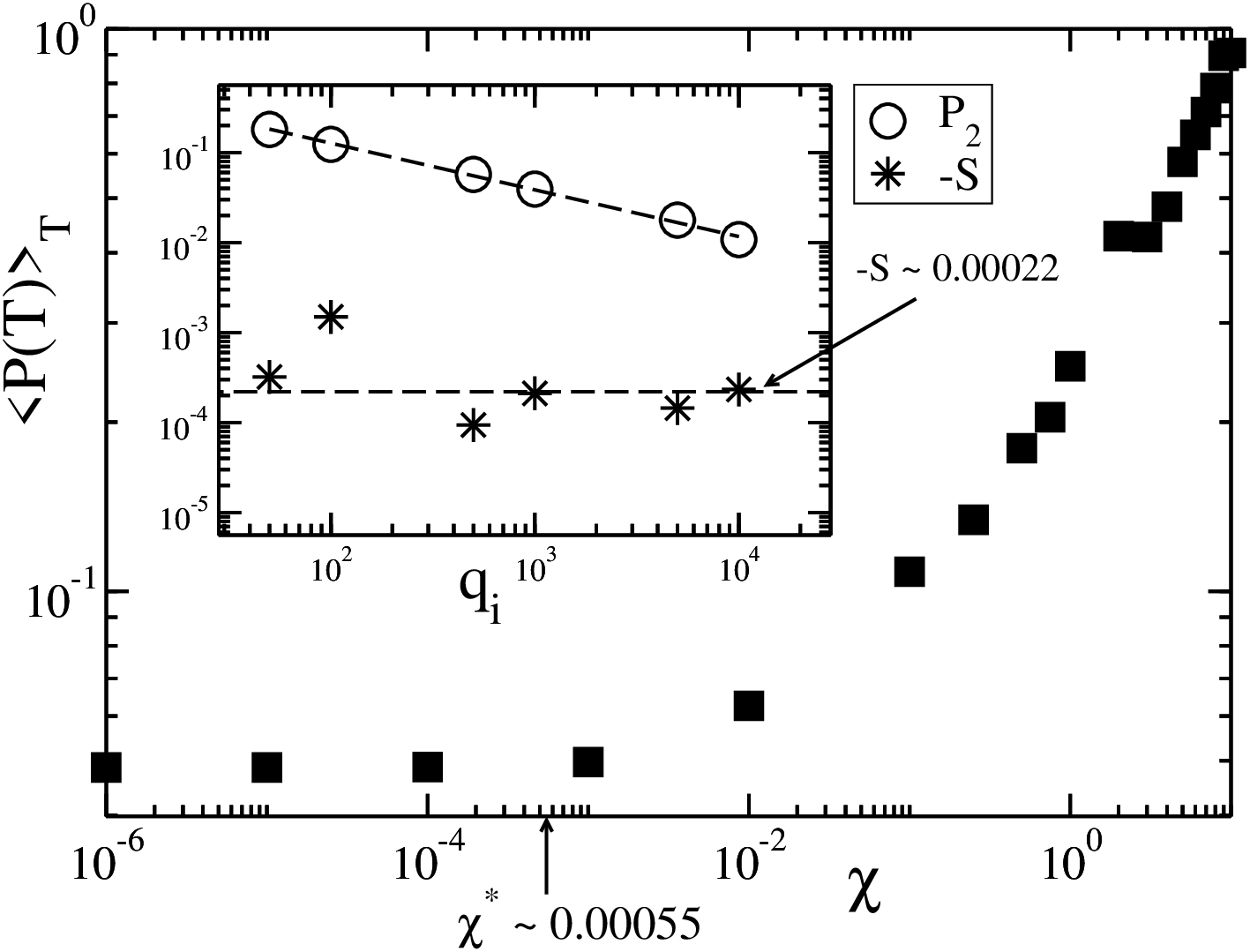}
\caption{\label{cap:fig4}
The integrated survival probability $\langle P(T) \rangle_{T}$ vs. $\chi$. For $\chi\geq 5.5\times 10^{-4}$ we observe
an increase in $\langle P(T) \rangle_{T}$. The first point in the curve corresponds to the case $\chi=0$
and is included in the log-log plot as a reference point.
Upper inset shows the scaling of the $P_2$ and $-S$ vs. the system size
$q_i$. The horizontal dashed line is $S = -2.2\times 10^{-4}$.
}\end{figure}

Going back to Fig.~\ref{cap:fig1}, we observe that the destruction of the anomalous diffusion of the linear model takes place for $\chi \ge \chi^*$.
To quantify $\chi^*$, we evaluate the time-
average survival probability $\langle P(T) \rangle_{T}$, defined as \cite{BDF93,T94,DK96}
\begin{equation}
\label{PT}
\langle P(T)\rangle_{T}\equiv \lim_{T\rightarrow\infty}{1\over T} \int_0^T |\psi_{n_0}(t)|^2 dt
\end{equation}
for various values of the nonlinearity strength $\chi$. In our numerical calculations, we took an average over a time interval of $T=20000$. Our numerical results are reported in Fig.~\ref{cap:fig4}. We see that up to $\chi^*
\approx 5.5 \times 10^{-4}$, the time average survival probability $\langle P(T)\rangle_{T}$ remains unchanged. 
As the nonlinearity strength $\chi$ is increased further, a fraction of the excitation begins to 
localize at the initial site. As a result, the fraction of the excitation that can propagate 
is now effectively smaller, leading 
to smaller values of $\langle m_2(t)\rangle$ as $\chi$ is increased. We note that a similar 
type of self-trapping phenomenon \cite{KC85,BDF93} was observed in various nonlinear lattices, 
\cite{LJR94,JHR95,PYFE92,DK96,PXG97,DJ98,AK93,T94} though in all these cases the value of 
$\chi^*$ was much larger, i.e. $\chi^*\sim O(1)$.

The following heuristic argument provides some understanding of the appearance of self-trapping 
phenomenon for the NLHM. We consider successive rational approximants $\sigma_i=p_i/q_i$ 
of the continued fraction expansion of $\sigma$. On a length scale $q_i$, the periodicity of the 
potential is not manifested and we may assume that for $\chi<\chi^*$ the eigenfunctions preserve their critical structure as in the case of $\chi=0$. 
In this case, the partitioning of the energy over the $q_i$ sites is 
\cite{hamilt}
\begin{equation}
\label{Hdnls}
{\cal H}=-\frac{\chi}{2}\sum_{n=1}^{q_i}\abs{\psi_n}^4
  + \sum_{n=1}^{q_i} (\psi_n^*\psi_{n-1} + \psi_{n-1}^*\psi_n)
  + \sum_{n=1}^{q_i} V_n\abs{\psi_n}^2
\end{equation}
Let us first estimate the energy ${\cal H}_{\rm ss}$ associated with the initial state $
\delta_{n,n_0}$. Since the probability distribution is located at a single site, we get an 
energy
\begin{equation}
\langle {\cal H}_{\rm ss}\rangle_{\phi}= -\frac{\chi}{2}
\end{equation}

Now we want to evaluate the partitioning of the energy ${\cal H}_{\rm ext}$ for a fractal wavefunction. 
In this case, the first term in Eq.~(\ref{Hdnls}) is the inverse participation number, which scales with the system size $q_i$ as
\begin{equation} 
P_2\sim q_i^{-D_2^{\psi}}
\end{equation}
Hence, in the thermodynamic limit where $q_i \rightarrow \infty $, this term will go to zero.
We define the sum of the two other terms as follows:
\begin{equation}
S= \langle \sum_{n=1}^{q_i} (\psi_n^*\psi_{n-1} + \psi_{n-1}^*\psi_n+V_n\abs{\psi_n}^2 )\rangle_{\phi}
\end{equation}
We found from numerical calculations (see inset of Fig.~\ref{cap:fig4}) that $S$ goes to a finite value in the thermodynamic limit, and, therefore, arrive at the equation $\langle {\cal H}_{\rm ext} \rangle_{\phi}=S$.

If ${\cal H}_{\rm ext}$ is higher than the initial energy $H_{\rm ss}$, then conservation of energy prevents the partitioning of energy 
over all $q_i$ sites. Therefore, in the thermodymanic limit, we get for the critical nonlinearity $\chi^*$ that 
\begin{equation}
\chi^*\sim -2 S.
\end{equation}
The numerical data in Fig.~\ref{cap:fig4}~(inset) indicate that $S\sim -2.2\times 10^{-4}$, which is consistent with our numerical evaluation of $\chi^*\sim 5.5\times 10^{-4}$.

%

\section{Conclusions}
In this paper, we studied the nonlinear Harper model at criticality and found bounds for the power-law exponent of the temporal spreading of the wavepacket variance.
These bounds reflect the fractal dimension of the LDoS of the linear system.
This nonlinear spreading appears for nonlinearity strength above some value and persists up to time $t^{*} \sim \chi^{1/D_2^{\mu}}$, which depends parametrically on the nonlinearity strength. After this time, the linear spreading of the second moment is restored; other moments, however, are still affected by the
nonlinearity. For the
central part of the evolving profile, we have also found a scaling
relation that applies to
any time and any nonlinearity strength.


The importance of these results lies in their applicability to quasiperiodic photonic structures (such 
as optical super-lattices\cite{KHM90}) and arrays of magnetic micro-traps for atomic Bose-Einstein 
Condensates\cite{JZ03} (BEC). In the latter case, nonlinear wave self-action appears due
to atom-atom scattering in BEC, while in optical crystals, it appears due to
light-matter interactions.

We would like to acknowledge T. Geisel for his continuous interest and support of this
project as well as D. Cohen and S. Wimberger for their contributions to useful discussions.



\begin{thebibliography}{99}
\bibitem{note1} Our numerical simulations showed that, in general, $\delta n = 1$ or $2$ sites.
\bibitem{hamilt} We recall that the DNLS (\ref{dnls}) can be derived from the Hamiltonian
Eq.~(\ref{Hdnls}), using the Poisson brackets $\{\psi_m,\psi_n^*\}=i\delta_{mn}$, $\{\psi_m,
\psi_n\}=\{\psi_m^*,\psi_n^*\}=0$, and the equation of motion ${\dot \psi_n}=\{H,\psi_n\}$.
\end{thebibliography}
\end{document}